\documentclass[twocolumn,nodate,showpacs,preprintnumbers,nofootinbib,amsmath,amssymb,aps,prd]{revtex4}
\usepackage{graphicx,verbatim}
\usepackage{amsmath}
\usepackage{amsfonts}
\usepackage{amssymb}
\usepackage{stmaryrd}
\usepackage{color}
\usepackage{float}

\def\scri{\mathcal{I}}
\def\scrip{\mathcal{I}^{+}}
\def\rmd{\mathrm{d}}
\def\={\hat{=}}

\def\vk{\vec{k}}
\def\e{\mathfrak{e}}

\def\E{\mathcal{E}}
\def\B{\mathcal{B}}

\def\etar{\eta_{\text{\tiny{ret}}}}

\def\Qp{Q^{(p)}}

\def\ord{\mathcal{O}}

\def\Diff{{\rm Diff}(\mathcal{I})}
\def\Diffp{{\rm Diff}(\mathcal{I}^{+})}
\def\BMS{\mathfrak{B}}
\def\Poin{\mathcal{P}}

\def\h{\hat}

\def\be{\begin{equation}}
\def\ee{\end{equation}}
\def\ba{\begin{eqnarray}}
\def\ea{\end{eqnarray}}
\def\f{\frac}

\def\Gds{\mathcal{G}_{\rm dS}}
\def\qo{\mathring{q}}
\def\go{\mathring{g}}

\def\vx{\vec{x}}

\begin{document}

\title{Gravitational waves from isolated systems:\\ 
Surprising consequences of a positive cosmological constant}

\author{Abhay Ashtekar}
\email{ashtekar@gravity.psu.edu} \affiliation{Institute for
Gravitation and the Cosmos \& Physics
  Department, Penn State, University Park, PA 16802, U.S.A.}
\author{B\'eatrice Bonga}
\email{bpb165@psu.edu} \affiliation{Institute for Gravitation and
the Cosmos \& Physics
  Department, Penn State, University Park, PA 16802, U.S.A.}
\author{Aruna Kesavan}
\email{aok5232@psu.edu} \affiliation{Institute for Gravitation and the
Cosmos \& Physics Department, Penn State, University Park, PA 16802,
U.S.A.}

\begin{abstract}

There is a deep tension between the well-developed theory of gravitational waves from isolated systems and the presence of a positive cosmological constant $\Lambda$, however tiny. In particular, 
even the post-Newtonian quadrupole formula, derived by Einstein in 1918, has not been generalized to include a positive $\Lambda$. We first explain the principal difficulties and then show that it is possible to overcome them in the weak field limit. These results also provide concrete hints for constructing the $\Lambda >0$ generalization of the Bondi-Sachs framework for full, non-linear general relativity.  

\end{abstract}


\maketitle

Although Einstein showed that general relativity admits gravitational waves in the \emph{weak field approximation} already in 1916, in 1936 he suggested that this result is an artifact of linearization \cite{Kennefick1}! Confusion on the reality  of gravitational waves in \emph{full} non-linear theory persisted until the 1960s \cite{Kennefick2}. It was finally dispelled through the work of Bondi, Sachs and others \cite{bondi}. Penrose geometrized this framework by introducing the notion of a conformal completion whose boundary, $\scri$, serves as the natural arena to analyze gravitational radiation \cite{rp}. There is a coordinate invariant notion at $\scri$, now called the \emph{Bondi news tensor} \cite{aa-radmodes}, that characterizes the presence of gravitational waves. Expressions \cite{bondi,aams} of energy, momentum and angular momentum carried away by gravitational waves across the future boundary $\scrip$  are widely used to extract astrophysical effects --such as `black hole kicks' \cite{kicks}-- from numerical simulations. Once the conceptual issues are thus resolved, geometric optics approximation can also be used to extract physical properties of radiation in the `wave zone', without direct reference to $\scri$ \cite{go}.

This well-developed framework assumes Einstein's equation with $\Lambda$=$0$. However, over the last two decades, observations have established that the universe is undergoing an accelerated expansion that is well-modeled by a positive $\Lambda$. It is then natural to inquire if the Bondi-Sachs theory can be extended to incorporate this feature. This task has turned out to be surprisingly difficult. For example, the Weyl tensor component $\Psi_{4}^{0}$, routinely used to construct wave forms in numerical simulations in the $\Lambda=0$ case, acquires ambiguities even at $\scrip$ if $\Lambda>0$  \cite{bicaketal,rp,rp2}. Indeed, we do not yet have the analog of the Bondi news to characterize gravitational waves in a gauge invariant manner, nor expressions of energy and momentum they carry. Unforeseen difficulties arise already in the weak field limit. Consequently, even Einstein's 1918 quadrupole formula \cite{Einstein:1918} is yet to be generalized. Given the early confusion on the reality of gravitational waves, it is important to improve on this situation.

The goal of this letter is to first succinctly summarize the obstacles 
and then show that they can be overcome in the weak field limit using strategies that are well suited for the full theory. In particular, we will present the desired generalization of Einstein's quadrupole formula.

\emph{Full, non-linear theory.} Because gravitational waves are ripples on the very geometry of space-time, they lead to a surprising feature already in the $\Lambda$=0 case. While every asymptotically flat metric $\h{g}_{ab}$ approaches a Minkowski metric $\h{\eta}_{ab}$ near $\scri$ (up to $\mathcal{O}(1/r)$ terms), the presence of gravitational waves introduces an essential ambiguity in the choice of $\h{\eta}_{ab}$. Consequently, contrary to one's first expectations, the asymptotic symmetry group is not the Poincar\'e group $\Poin$ but the infinite dimensional Bondi-Metzner-Sachs (BMS) group $\BMS$, and reduces to $\Poin$ only when the Bondi news vanishes \cite{aa-radmodes,abk1}. The effect of gravitational waves is even more striking for $\Lambda>0$ \cite{abk1}: As we now explain, the physical metric $\h{g}_{ab}$ differs from the de Sitter metric even to leading order (albeit in a controlled manner).
 
The precise definition \cite{rp,abk1} of asymptotically de Sitter space-times mimics that of asymptotically flat space-times, except that Einstein's equation now has a positive $\Lambda$ term. The asymptotic conditions are chosen to accommodate familiar examples such as the Kerr-de Sitter space-time and, in spite of their spatial homogeneity, the Friedmann-Lema\^itre-Robertson-Walker (FLRW) models with `dark energy'. However, unlike in the $\Lambda=0$ case, we no longer have a unique conformal class of intrinsic metrics on $\scri$. As a result, now the asymptotic symmetry group is the entire $\Diff$ \cite{strominger,abk1} which does not admit a preferred 4-dimensional group of translations. Therefore, one cannot even begin to introduce the notion of energy-momentum  carried by gravitational waves. 

One's first reaction would be to strengthen the boundary conditions to suitably reduce $\Diff$. A natural strategy is to require that the intrinsic, positive definite metric $q_{ab}$ at $\scri$ be conformally flat \emph{as it is in de Sitter space-time.} Familiar examples satisfy this additional condition, and it immediately reduces $\Diff$ to the 
10-dimensional group of conformal isometries of $q_{ab}$ which, furthermore, is naturally isomorphic to the de Sitter group $\Gds$. Therefore, we can hope to define de Sitter momenta. Indeed, using the field equation, a systematic procedure leads one to associate a de Sitter charge $Q_{\xi}[C] = \oint_{C} \E_{ab}\xi^{a} \rmd S^{b}$ with any 2-sphere cross-section $C$ of $\scri$ and any generator $\xi^{a}$ of $\Gds$, where $\E_{ab}$ is the electric part of the leading-order Weyl tensor \cite{abk1}. In Kerr-de Sitter space-time, the only non-vanishing de Sitter charges are the (correctly normalized) mass and angular momentum. Thus, at first glance, the strategy of strengthening the boundary conditions appears to be successful.

However, a detailed calculation shows that the field equation and Bianchi identities imply that requiring conformal flatness of $q_{ab}$ is \emph{equivalent to} demanding that the magnetic part $\B_{ab}$ of the leading order Weyl curvature should vanish at $\scri$. Since $\scri$ is space-like, it is clear from Friedrich's pioneering work \cite{hf} that this is a severe mathematical restriction. The weak field analysis reported below brings out the physical meaning of this restriction explicitly. Furthermore it turns out that, in absence of matter fields at $\scri$, the charges $Q_{\xi}[C]$ are absolutely conserved; if $q_{ab}$ is required to be conformally flat, gravitational waves do not carry any de Sitter momenta! Thus, we are faced with a quandary: While requiring conformal flatness of $q_{ab}$ seems physically unreasonable, dropping it leaves us with too weak a structure to discuss the physics of gravitational waves. 

\emph{Remark:}\, In terms of the physical metric $\h{g}_{ab}$, the situation can be summarized as follows. First, the Bondi ansatz has to be suitably generalized to accommodate a positive $\Lambda$. One can then solve Einstein's equations asymptotically \cite{hc} to obtain the metric $q_{ab}$ on $\scrip$, following \cite{abk1}. In the {axi-symmetric case}, one obtains 
\be \label{hc} q_{ab}\, \rmd x^{a} \rmd x^{b} = \rmd u^{2} +e^{2\Lambda f}\, \rm{d}\theta^{2} + e^{-2\Lambda f}\, \sin^{2}\theta\,\rm{d}\varphi^{2} \ee
with $f$=$f(u,\theta)$.\!\! The single radiative mode (or polari\-zation) of axisymmetric gravitational waves is encoded in the `shear-term' via $f$. The 3-metric $q_{ab}$ is conformally flat if and only if $f$=0, i.e., the radiative mode is absent. Thus,  if we allow gravitational radiation near $\scrip$, then the physical metric differs from the de Sitter metric already \emph{at leading order} in the Bondi-type expansion!\\

\emph{Linearization off de Sitter space-time:}  As in the $\Lambda=0$ case, analysis of linearized gravitational waves provides intuition and guidance for the development of the full non-linear theory. Since
the background de Sitter metric naturally reduces $\Diff$ to $\Gds$, the quandary in the non-linear case is simply bypassed in this approximation. But as we now discuss, other complications remain.
  
As in the cosmology literature, we will restrict ourselves to the upper Poincar\'e patch of de Sitter space-time (although the linearized solutions we obtain can be extended to full de Sitter space-time using Cauchy evolution). It is convenient to express the perturbed physical metric $\h{g}_{ab}$ using conformal time $\eta$ as 
\be \h{g}_{ab} = a^{2}(\eta)\, (\go_{ab} + \epsilon h_{ab})\quad {\rm with}\quad \go_{ab}\rmd x^{a} \rmd x^{b} = -\rmd \eta^{2} + \rmd {\vec x}^{2} \nonumber \ee 
where $a(\eta) = -1/H\eta \equiv -\sqrt{3}/(\sqrt{\Lambda}\,\eta)$ is the de Sitter scale factor, and $\epsilon$, the (mathematical) smallness parameter. Then, in the transverse traceless gauge, the general solution to linearized Einstein's equation is given by
\be \label{ft} h_{ab}(\vx,\,\eta) \, \equiv\, \int \f{\rmd^{3}k}{(2\pi)^{3}}\, \sum_{(s)=1}^{2}\, \, h_{\vk}^{(s)} (\eta)\, \e^{(s)}_{ab}(\vk)\,\, e^{i\vk\cdot\vx}\, ,\ee
where $(s)$ labels the two helicity states, $\e^{(s)}_{ab}(\vk)$ are the standard polarization tensors, and 
\ba \label{sol1} h_{\vk}^{(s)}(\eta) = (-2H)&\!\!\!\!\!\big[ E^{(s)}_{\vk}\, (\eta\, \cos (k\eta) - (1/k)\,\sin (k\eta))\nonumber\\  -& \!\! B^{(s)}_{\vk} \, (\eta\sin (k\eta) + (1/k)\, \cos (k\eta)) \big]\, . \ea
Here $E^{(s)}_{\vk}$ and $B^{(s)}_{\vk}$ are arbitrary coefficients, determined by the initial data of the solution. In a standard completion of de Sitter space-time, $\scrip$  is the $\eta$=0 surface and $\go_{ab}$ serves as the conformally rescaled metric which is well-behaved at $\scrip$. By inspection, the mode with coefficient $E^{(s)}_{\vk}$ vanishes at $\scrip$ and is called the `decaying mode' in the cosmology literature while the mode with the coefficient $B^{(s)}_{\vk}$ is non-zero at $\scrip$ and is called the `growing mode'. 

\begin{figure}[]
  \begin{center}
  \vskip-0.4cm
        \includegraphics[width=2.3in,height=2in,angle=0]{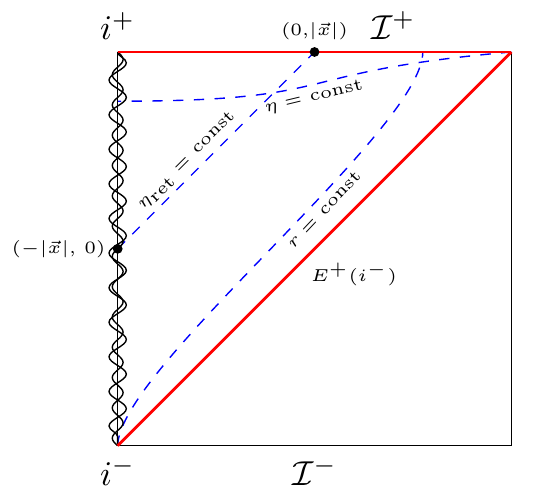}
\caption{ The rate of change of quadrupole moments at the point $(-|\vec{x}|, \vec{0})$ on the source creates the retarded field at the point $(0, \vec{x})$ on $\scrip$. The figure also shows the cosmological foliation $\eta= {\rm const}$ and the time-like surfaces $r={\rm const}$. As $r$ goes to infinity, the $r:= |\vx| ={\rm const}$ surfaces approach $E^{+}(i^{-})$. Therefore, in contrast with the situation in Minkowski space-time, for sufficiently large values of $r$, there is no flux of energy across the $r={\rm const}$ surfaces.}
\label{poincare}
\end{center}
\end{figure} 

Let us now go beyond the cosmology literature by calculating de Sitter momenta carried by gravitational waves. The covariant phase space framework provides a natural avenue that avoids the use of pseudo-tensors which caused much confusion in the early days on whether gravitational waves carry energy \cite{Kennefick2}. Of particular interest is the expression $E_{T}$ of energy associated with a de Sitter time-translation $T = -H (\eta\,\partial_{\eta} + \vx \cdot\partial_{\vx})$. It is given by \cite{abk2}:
\be \label{E} E_{T} = \f{{H}}{8\pi G} \int\!\! \f{\rmd^{3}k}{(2\pi)^{3}}\, k \,\,{\sum_{(s)=1}^{2}}\, E^{(s)}_{\vk} \mathcal{L}_{\vk} (B^{(s)}_{\vk})^{\star} + 2\, E^{(s)}_{\vk} \, (B^{(s)}_{\vk})^{\star}. \ee
Note that $\scrip$ is space-like and since all Killing vectors, including de Sitter time translations $T^{a}$, must be tangential to $\scrip$, they are space-like in some neighborhood of $\scrip$. Therefore, one would expect the Hamiltonian generating the canonical transformation induced by $T^{a}$ not to have a definite sign. This expectation is explicitly borne out in Eq.\,(\ref{E}): The de Sitter energy $E_{T}$ can be negative and with arbitrarily large magnitude, no matter how small $\Lambda$ is. The limit $\Lambda\to 0$ requires care; one has to use the differential structure defined by $(t,\vx )$ rather than $(\eta, \vx)$ where $t$ is the proper time, given by $e^{-Ht} = - H\eta$. When this is done, the de Sitter time translation $T^{a}$ goes over to a Minkowski time translation and $E_{T}$ to energy in Minkowski space-time, which is of course positive. Thus, there is a surprise: gravitational (and also electromagnetic) \emph{waves can carry negative energy} $E_{T}$ in de Sitter space, and the lower bound on $E_{T}$ is \emph{infinitely discontinuous} in the limit $\Lambda \to 0$. 

Finally, one can show that the magnetic part $\B_{ab}$ of the leading order, linearized Weyl curvature vanishes at $\scrip$ \emph{if and only if} the coefficients $B^{(s)}_{\vk}$ are all identically zero. Thus, if we demand that perturbation $h_{ab}(\vx,\eta)$ should respect the conformal flatness of the de Sitter metric to first order, we eliminate all the `growing modes', cutting by fiat the space of allowed linearized perturbations by half. Furthermore, while the decaying modes do remain, (\ref{E}) implies that the conserved energy $E_{T}$ carried by them vanishes identically. (Same is true for linear and angular momentum.) This discussion brings out that conformal flatness condition on $q_{ab}$ is a \emph{physically} inadmissible restriction in the discussion of gravitational waves.\\   

\emph{The quadrupole formula:} Consider a time varying quadrupole in de Sitter space-time, depicted in Fig.~\ref{poincare}. Our task is to calculate the flux of energy carried by gravitational waves across $\scrip$ in the post-de Sitter, first post-Newtonian approximation. Surprisingly, one faces several new difficulties. We will give four illustrative examples. First, since gravitational waves can carry arbitrarily large negative energy in de Sitter space-time, a priori, there is a danger of a gravitational instability. Second, in the standard $\Lambda$=0 derivation, one considers\, $r$=const\, time-like cylinders with large $r$ which approach $\scrip$, and makes heavy use of $1/r$-expansions to calculate the energy flux through these cylinders. In particular, the `transverse-traceless decomposition' used  in this calculation is gauge invariant only up to $\mathcal{O}(1/r^{2})$ terms (see, e.g. \cite{pw}). In de Sitter space-time, on the other hand, such time-like cylinders approach the past cosmological horizon $E^{+}(i^{-})$ rather than $\scrip$ and the energy flux across $E^{+}(i^{-})$ vanishes for retarded solutions. Therefore, a new approximation scheme tailored to the de Sitter $\scrip$ is needed. Third, in contrast to the $\Lambda$=0 case, the propagation of the metric perturbation $h_{ab}$ is not sharp: there is also a tail term which is as significant near $\scrip$ as the sharp term, creating the possibility that Einstein's quadrupole formula may be significantly modified.  Finally, the physical wavelengths $\lambda$ of perturbations grow as the waves propagate and can vastly exceed the curvature radius in the asymptotic region,  making the standard $\lambda\ll r$ expansion (and the geometric optics approximation) untenable near $\scrip$. Because of these new features, it is not clear that one's initial expectation that a tiny cosmological constant will only modify Einstein's quadrupole formula negligibly is correct. Indeed, as we saw, physical quantities can be discontinuous at $\Lambda$=0. However, our detailed investigation shows that several factors intervene to alleviate these apparently menacing implications of a positive $\Lambda$ for sources of interest to the current gravitational wave observatories \cite{abk3}. 

The first new element is to replace the $1/r$-expansion used in the $\Lambda>0$ case with a well-controlled `late time' expansion; i.e. to approach $\scrip$ using\, $\eta$=const\, \emph{space-like} surfaces. Consider retarded (trace-reversed) solutions $h_{ab}$ to linearized Einstein's equation on de Sitter space-time sourced by a first order stress-energy tensor $T_{ab}$ depicted in Fig.~\ref{poincare}, given in \cite{vrs}. We assume that the physical size $D(\eta)$ of the source is uniformly bounded by some $D_{o} \ll \ell_{\Lambda}$ where $\ell_{\Lambda} = \sqrt{3/\Lambda}$ is the cosmological radius. We can restrict ourselves to the upper Poincar\'e patch of de Sitter space-time because observers in the lower Poincar\'e patch cannot see the source nor detect the emitted radiation. One can extract the leading order piece of this post-de Sitter solution using the late time and first post-Newtonian approximations, i.e., ignoring $\ord\big((D_{o}/\ell_{\Lambda})(1-\eta/r)\big)$ and $\ord(v/c)$ terms relative to those that are $\ord(1)$. The solution, of course, involves integrals over the stress-energy tensor. 

In a second step, using the continuity equation, one can replace these integrals with source quadrupoles following the procedure used in the $\Lambda$=0 case. The mass (or, more precisely, the density) quadrupole moment is defined in an invariant manner as an integral on each\, $\eta$=const\, slice:
\be \label{Q} Q_{ab}^{(\rho)}(\eta) = {\int}\! \rmd^{3}V\,\, \rho(\eta)\, (ax_{a})(ax_{b})\, . \ee
However, one finds that just as pressure contributes to gravitational attraction in the Raychaudhuri equation in cosmology, now a time changing \emph{pressure} quadrupole $Q_{ab}^{(p)}$ (obtained by replacing $\rho(\eta)$ with $p(\eta)$ in (\ref{Q})) also contributes. The expression of $h_{ab}(\eta,\vx)$ has a sharp propagation term that is sensitive only to the time derivatives of the quadrupoles at the retarded instant $\etar$, as well as a tail term involving an integral over $\eta\in (-\infty, \etar)$ which is absent in the $\Lambda$=0 case. 

In the final step, we can use the covariant phase space framework \cite{abk2} to extract the energy $E_{T}$ carried by the linearized gravitational wave $h_{ab}$, where $T^{a}$ is the de Sitter Killing vector tailored to the source via $i^{\pm}$. We find:
\be
\label{energy}
E_T\, \hat{=}\, \frac{G}{8 \pi} \int_{\scrip}\!\! {\rm d} u\, {\rm d}^{2}S\, \Big[\mathcal{R}^{ab}(\vx)\, \mathcal{R}_{ab}^{TT}(\vx)\,\Big]\, .\ee
Here, $u$ is the affine parameter of $T^{a}$; the `radiation field' $\mathcal{R}_{ab}(\vx)$ on $\scrip$ is given by 
\ba\label{R} \mathcal{R}_{ab}(\vx) &\hat{=}& \Big[\dddot{Q}_{ab}^{(\rho)} +3 H \ddot{Q}_{ab}^{(\rho)} + 2H^2 \dot{Q}^{(\rho)}_{ab} \nonumber\\
&+& H  \ddot{Q}_{ab}^{(p)} + 3H^2  \dot{Q}_{ab}^{(p)} + 2 H^3 \Qp_{ab}\Big](\etar)\, ; \ea
and $\mathcal{R}_{ab}^{TT}(\vx)$ is its divergence-free and trace-free part  on $\scrip$. The `dot' stands for Lie-derivative with respect to the Killing field $T^{a}$. (For details, see \cite{abk3}.)

The radiated energy has the following properties:\\ (i) one can show that $E_{T}$ is positive definite although $T^{a}$, being tangential to $\scrip$, is space-like; \\(ii) $E_{T}$ vanishes identically if the source is stationary, i.e., $\mathcal{L}_{T} T_{ab} =0$ despite the fact that there is a term $\Qp_{ab}$ without a `dot'; \\(iii) Since $H\to 0$ in the limit $\Lambda \to 0$, we recover the Einstein quadrupole formula in this limit; and,\\ (iv) If the dynamical time scale associated with the source is negligible compared to the Hubble time, then $\mathcal{R}_{ab} \approx \dddot{Q}_{ab}^{(\rho)}$, whence Einstein's quadrupole formula provides an excellent approximation to the energy loss for compact binaries of interest to LIGO observatories.

`Tameness' of these properties raises the question: Why did the menace of a positive $\Lambda$ turn out to be a phantom for sources of current interest? First, all gravitational waves with negative de Sitter energy have non-trivial fluxes across the past cosmological horizon $E^{+}(i^{-})$. Since these fluxes vanish for retarded solutions, negative energy solutions cannot arise from time changing quadrupoles. Second, we replaced the $1/r$-expansions and the approximate notion of `transverse-traceless fields' used in the $\Lambda$=0 analysis \cite{pw} with a late-time expansion and the exactly gauge invariant notion of transverse-traceless fields (normally used in cosmology). Third, while the propagation of the metric perturbation $h_{ab}$ is not sharp, that of its time derivative turns out to be sharp and it is only the time derivative that features in the expression of radiated energy. Finally, although the wavelengths of perturbations do grow as they propagate, because the time derivatives in (\ref{energy}) are evaluated \emph{at retarded time} $\etar$, what matters is the wavelength at the emission time, rather than at late times near $\scrip$.\\

\emph{Summary and outlook:} Given the early confusion on whether gravitational waves are physical and the fact that the gravitational wave observatories are now on the threshold of opening a new window on the universe, it is important that the theoretical foundations of the subject be solid. A priori, one cannot be certain that the effects of $\Lambda$ would be necessarily negligible because, irrespective of how small its value is, its mere presence introduces several conceptual complications requiring a significant revision of the standard framework. Our analysis makes the errors involved in setting $\Lambda=0$ explicit. In particular, in the post-de Sitter, first post-Newtonian approximation, it shows that these complications are harmless for binary systems that are the primary targets of the current observatories. But some of the subtleties associated with a non-zero $\Lambda$ could be important for future detectors such as the Einstein Telescope that will receive signals from well beyond the cosmological radius. They may also be important for the analysis of the very long wave length radiation produced by the first black holes. To probe such issues, the current analysis is being extended to FLRW space-times with positive $\Lambda$.

Our results also provide guidance for  constructing the $\Lambda >0$ extension of the Bondi-Sachs framework for the \emph{full, non-linear} theory. First, even in the absence of a clear-cut candidate replacing the Bondi news tensor, one could impose the `no incoming radiation' condition by asking that there be no flux of fields propagating into space-time across the past cosmological horizon $E^{+}(i^{-})$, a condition that can be neatly captured by requiring that $E^{+}(i^{-})$ be a weakly isolated horizon \cite{wih2}. Second, in the linear theory we used the background de Sitter metric to reduce $\Diffp$ to $\Gds$. In the non-linear case, given the intrinsic metric $q_{ab}$ at $\scrip$ one can attempt to extract from it a conformally flat metric $\qo_{ab}$ in an invariant manner, e.g., by setting $f$=0 in Eq.(\ref{hc}). The asymptotic symmetries $\xi^{a}$ would then be conformal Killing fields of this `background metric' $\qo_{ab}$ and de Sitter fluxes \cite{abk1} would not vanish because $\xi^{a}$ are not conformal Killing fields of $q_{ab}$. Finally, in the $\Lambda$=0 case, fluxes of the BMS momenta across $\scrip$ \cite{aams} are closely related to their linear analogs. Our linear flux expressions already provide strong hints on the flux expressions in the full theory.

\section*{Acknowledgments} This work was supported in part by the NSF grant PHY-1505411. We would like to thank Luis Lehner, Eric Poisson, B. Sathyaprakash and Ray Weiss for discussions.


\begin{thebibliography}{99}

\bibitem{Kennefick1} D.~Kennefick, Controversies in the history of the radiation reaction problem in general relativity, \texttt{arXiv:9704002}.

\bibitem{Kennefick2} D.~Kennefick, Traveling at the speed of thought: Einstein and the quest for gravitational waves, (Princeton university press, Princeton (2007)).

\bibitem{bondi} H.~Bondi, M.~van der Burg, and A.~Metzner, Proc. R. Soc. (London) A\textbf{269}, 21 (1962); R.~K.~Sachs, Proc. R. Soc. (London) A \textbf{270}, 103 (1962).

\bibitem{rp} R.~Penrose, Proc. R. Soc. (London) A\textbf{284}, 159-203 (1965).

\bibitem{aa-radmodes}A.~Ashtekar, J. Math. Phys. \textbf{22}, 2885-2895 (1981); In: \emph{Surveys in Differential Geometry}, edited by L.~Bierri and S.~T.~Yau (International Press, Boston, 2015).

\bibitem{aams} A.~Ashtekar and M.~Streubel, Proc. R. Soc. (London) A \textbf{376}, 585-607 (1981).

\bibitem{kicks}M.~Campanelli, C.~O.~Lousto, Y.~Zlochower, D.~Merritt,
Phys. Rev. Lett. \textbf{98}, 231102 (2007);\\ 
J.~G.~Baker, W.~D.~Boggs, J.~Centrella, B.~J.~Kelly, S.~T.~McWilliams, M.~C.~Miller and J.~R.`van Meter,  Astrophys. J. \textbf{682}, L29-L32 (2008). 

\bibitem{go} K.~S.~Thorne, In: \emph{Gravitational Radiation}, edited by N.~Deruelle and S.~Piran, (North-Holland Amsterdam, 1983).


\bibitem{bicaketal} P.~Krtou\v{s} and J.~Podolsk\'y, Class. Quantum Grav. \textbf{21}, R233-R273 (2004).
  
\bibitem{rp2} R.~Penrose, Gen. Rel. Grav. \textbf{43}, 3355-3366 (2011). 

\bibitem{Einstein:1918}A.~Einstein, Sitzungsberichte der K\"{o}niglich Preussischen Akademie der Wissenschaften, Berlin, (1918). 

\bibitem{strominger} D.~Anninos, G.~S.~Ng, A.~Strominger, Quant. Grav. {\bf 28} 175019 (2011).
 
\bibitem{abk1} A.~Ashtekar, B.~Bonga and A.~Kesavan, Class. Quant. Grav. \textbf{32}, 025004 (2015). 

\bibitem{hf} H.~Friedrich, J. Diff. Geo. \textbf{3}, 101-117 (1986).

\bibitem{hc}X.~He and Z.~Cao, Int. J. Mod. Phys. \textbf{D}24, 1550081 (2015).

\bibitem{abk2} A.~Ashtekar, B.~Bonga and A.~Kesavan, Phys. Rev. D \textbf{92}, 044011 (2015).

\bibitem{pw} E.~Poisson and C.~Will, \emph{Gravity: Newtonian, Post-Newtonian, Relativistic} (Cambridge University Press, Cambridge 2005).  

\bibitem{abk3}A.~Ashtekar, B.~Bonga and A.~Kesavan, Phys. Rev. D\textbf{92}, 10432(2015)  

\bibitem{vrs} H.~J.~de Vega, J.~Ramirez and N.~Sanchez, Phys. Rev. D \textbf{60} 044007 (1999).

\bibitem{wih2}A.~Ashtekar, C.~Beetle and J.~Lewandowski, Class. Quant. Grav. \textbf{19}, 1195-1225 (2002). 


\end{thebibliography}
\end{document}